\documentclass[11pt]{article}
\usepackage{coling2020}
\usepackage{times}
\usepackage{url}
\usepackage{latexsym}
\usepackage{xcolor}
\usepackage[ruled,vlined]{algorithm2e}
\usepackage{amsmath}
\usepackage[pdftex]{graphicx}
\usepackage{multirow}
\usepackage{subcaption}
\usepackage{wrapfig}

\colingfinalcopy 

\title{Large-scale Hybrid Approach for Predicting User Satisfaction with Conversational Agents}

\author{
  {\bf Dookun Park} \quad {\bf Hao Yuan} \quad {\bf Dongmin Kim} \quad {\bf Yinglei Zhang} \\
  {\bf Spyros Matsoukas} \quad {\bf Young-Bum Kim} \quad {\bf Ruhi Sarikaya} \quad {\bf Edward Guo}\\
  {\bf Yuan Ling} \quad {\bf Kevin Quinn} \quad {\bf Pham Hung} \quad {\bf Benjamin Yao} \quad {\bf Sungjin Lee}\\
  Alexa AI\\
  {\tt \{dkpark, yuanha, kdongmin, yingleiz, matsouka,}\\
  {\tt youngbum, rsarikay, guochenl, yualing, kevquinn,}\\
  {\tt hupha, benjamy, sungjinl\}@amazon.com} \\\AND
  }
  
\date{}

\begin{document}
\maketitle

\begin{abstract}
Measuring user satisfaction level is a challenging task, and a critical component in developing large-scale conversational agent systems serving the needs of real users. An widely used approach to tackle this is to collect human annotation data and use them for evaluation or modeling. Human annotation based approaches are easier to control, but hard to scale. A novel alternative approach is to collect user's direct feedback via a feedback elicitation system embedded to the conversational agent system, and use the collected user feedback to train a machine-learned model for generalization. User feedback is the best proxy for user satisfaction, but is not available for some ineligible intents and certain situations. Thus, these two types of approaches are complementary to each other. In this work, we tackle the user satisfaction assessment problem with a hybrid approach that fuses explicit user feedback, user satisfaction predictions inferred by two machine-learned models, one trained on user feedback data and the other human annotation data. The hybrid approach is based on a waterfall policy, and the experimental results with Amazon Alexa's large-scale datasets show significant improvements in inferring user satisfaction. A detailed hybrid architecture, an in-depth analysis on user feedback data, and an algorithm that generates data sets to properly simulate the live traffic are presented in this paper.

\end{abstract}

\section{Introduction}\label{introduction}
Developing intelligent conversational agents~\cite{McTear:02} is a topic of great interest in Artificial Intelligence, and there already are several large-scale conversational agents on the market, such as Amazon Alexa, Google Assistant, Apple Siri, and Microsoft Cortana, that have significant numbers of users and usages~\cite{conversational_agent_usage}\cite{Sarikaya_PDA:17}. 
Conversational agents today employ a diverse set of metrics to capture different aspects of business and user experience. For example, there are topline business metrics to track and drive success of the business: monthly active user, dialog count per user, downstream impact from highly valued action and negatively valued action. However, these metrics are normally not sensitive enough to detect changes in a timely fashion, requiring long experimentation times, and hence will result in slow experiment turnaround cycle. On the other hand, there are user experience metrics that capture unhandled user requests or dissatisfying user experiences due to system errors, incomplete service coverage, or poor response quality. User experience metrics generally have positive correlation with business metrics and tend to be fast moving and more sensitive than business metrics, and hence are suitable for experimentation to make data-driven decisions. Fine-grained user experience metrics are also a key to providing actionable insights to conversational skill developers by correlating metric shifts with interpretable factors such as user intent and slot.

In industry, manual annotation has been widely adopted to assess user satisfaction. Given a pair of user utterance and system response with its surrounding conversation history and contextual signals, via various mapping techniques including recent deep neural-net based approaches in Alexa~\cite{yb_Kim_alc:18}\cite{yb_Kim_naacl:18}, annotators return a user satisfaction score based on annotation guidelines. Due to its offline nature and limited bandwidth, however, it is ill-suited for online monitoring and experimentation or for providing actionable insights over a broad set of use cases. Although it is becoming common to build machine-learned models trained on manual annotations to mitigate such limitations, there still remain critical challenges: 1) Scalability: in Alexa today, there are tens of thousands skills available built by 3rd party public developers, and it is prohibitive to collect sufficient amounts of human annotations for all use cases; 2) Discrepancy with actual user satisfaction: annotators do not have full visibility into user's goal and context, and they make their best guess according to the annotation guidelines.

To address these challenges, in this paper, we explore the use of post experience user feedback. As shown in the example dialog below, we instrumented a feedback elicitation system to ask user's feedback with pre-designed prompts such as "Did I answer your question?" and to interpret user's response.
\begin{figure}[h]
    \begin{center}
        \framebox{
          \begin{minipage}{0.6\linewidth}
            \textbf{User:} When will it stop raining in New York?\\
            \textbf{Alexa:} In New York , intermittent rain is possible throughout the day. {Did I answer your question?}\\
            \textbf{User:} Yes.\\
            \textbf{Alexa:} Thanks for your feedback.
          \end{minipage}
        }
    \end{center}
	\caption{An example Alexa dialog that asks user feedback.}
	\label{fig:feedback_dialog}
\end{figure}

\noindent Unlike the conventional manual annotation, user feedback is a good proxy to user satisfaction as users know best whether Alexa provided the right experience they wanted, and the amount of user feedback can easily be scaled up to several orders of magnitude larger than that of human annotation. However, frequent feedback requests can introduce significant friction in user experience, thus we cautiously prompt users with a controlled rate. A machine-learned feedback prediction model is built to produce user satisfaction assessment when direct user feedback is unavailable.

During the course of our early exploration, we identified a few issues in collecting user feedback, which are preventing us from building a holistic metric using the user feedback signal only. Not all scenarios were applicable for collecting user feedback (e.g., when a user barges in and asks termination, asking a post-experience feedback can lead to an unnatural experience.), and not all domains/skills were able to onboard the feedback collection system at the same time. This caused an incomplete coverage of the user feedback data. 
In contrary, human annotation-based approaches are unobtrusive to user experience and applicable to all situations. 

Therefore, in this work, we propose a practical hybrid approach to take the best of both worlds. At a high level, our hybrid approach fuses direct user feedback and two types of predicted user satisfaction by two machine-learned models, one trained on user feedback data and the other on human annotation data. During the inference time, a waterfall policy is employed for each pair of user utterance and system response: 1) We first check if a direct user feedback is available, and respect it if available; 2) Otherwise, we check if the feedback-based prediction model is eligible and its prediction result shows a high confidence score. If it is, we take that feedback prediction; 3) Finally, in case we could not get a prediction from prior stages, we make a prediction with the human annotation-based model. On an Amazon Alexa's large-scale test dataset, our hybrid approach achieved significant improvements in precision, recall, F1-score, and PR-AUC (Precision-Recall Area Under Curve) by $4.4\%$, $28.7\%$, $18.3\%$, and $24.4\%$, respectively. Along with this performance improvement, in terms of data volume, the hybrid dataset became to have less dependency on human annotation as we were able to collect user feedback data collected at scale. This is another benefit of our hybrid approach.

The rest of the paper is structured as follows. In section~\ref{feedback_analysis}, we provide a brief analysis to understand the quality and traits of user feedback. In section~\ref{method}, we present our proposed hybrid approach. In section~\ref{expriment}, we describe our experimental setup and results. In section~\ref{related_work}, we summarize related work. Finally, in section~\ref{conclusion}, we conclude with discussion and future work.

\section{User feedback analysis}
\label{feedback_analysis}
This section provides an analysis to understand how user feedback correlates with human annotation. The primary dataset used contains 7,447 utterances with user feedback, and the dataset was annotated by trained annotators using the same annotation guidelines of our human annotation process. In the dataset, by the annotation work, about 35\% of utterances are mapped to other categories than YES or NO. Among these other categories, the biggest bucket is SILENCE where users did not provide any feedback. Our findings indicate that the majority of silence feedback correspond to satisfying experience. As a more in-depth analysis is required to fully understand other categories, in this work, we decided to utilize only those utterances with a YES or NO feedback which amounts to 4,729 utterances. By the human annotation work, the user feedback has an agreement rate with human annotation of 97.4\%, and a Cohen's kappa of 0.7877 (`substantial' agreement according to typical kappa interpretation~\footnote{\textit{Cohen's kappa coefficient}, \url{https://en.wikipedia.org/wiki/Cohen\%27s_kappa}}). Note that the high agreement rate between user feedback and human annotation is partly due to its bias toward satisfaction feedback as we do not have feedback elicitation opportunities when users barge in and ask for termination that are strongly correlated with user's dissatisfaction. To compensate this limitation, our hybrid approach supplement user feedback with human annotation when the use of user feedback is ineligible.

\section{Method}\label{method}
This section first describes a deep neural model that we designed to build predictive models for both user feedback and human annotation, and then provides the details of our hybrid approach to fuse several inputs of user satisfaction assessment. 

\subsection{Deep neural model for user satisfaction prediction}\label{dnn_model}
Before diving into modeling details, we first introduce a few terminologies. We define a pair of user utterance and Alexa response as a \textit{turn}. Figure~\ref{fig:feedback_dialog} shows an example dialog consisting of two turns, eliciting user feedback; We call the first turn the \textit{targeted turn} where Alexa asks "Did I answer your question?", and the second turn the \textit{answering turn}. 
Besides the user utterance and Alexa response, each turn also has some meta information, e.g., the timestamp of the turn, the conversational skill invoked to handle the turn, the active screen availability for the turn happened.~\footnote{Due to confidentiality reasons, we are not allowed to disclose the exact meta information features.} Thus, we represent a turn as 
\begin{equation}
    t_i = (u_i, a_i, f_i)
\end{equation}
where $u_i$ is user input text, $a_i$ is Alexa response text, and $f_i = [f_{i}^1, ..., f_i^k]$ is a list of meta information features. To capture contextual cues from the surrounding turns such as user's rephrasing patterns and barge-in patterns, we consider the dialog session including other turns around the targeted and answering turns. Rigorously, we define a \textit{session} as a maximal list of turns such that any two adjacent turns have a time gap less or equal to $\Delta$ minutes. (``Maximal'' means that if a session is a sub-list of some list of turns, then that list must have some adjacent two turns whose time gap is greater then $\Delta$.)
We denote a session as
\begin{equation}
    s_j = \langle t_i^j \rangle_{i=1}^{L_j}
\end{equation}
where $L_j$ represents the number of turns in session $j$.
Our dataset consisting of a set of session and label pairs can then be represented as
$$\{(s_j, \ y_j)\}_{j=1}^N$$
where $y_j$ denotes the binary user satisfaction with respect to the targeted turn in session $j$. 
Given session information $s_j$, the model produces a prediction score $p_j \in [0, 1]$ as the probability that the user is \textit{dissatisfied}. Note that we treat dissatisfaction as our primary class since accurately detecting defective experiences offers greater value for us to improve downstream components. 
Finally, there are some meta features are shared across turns in the same session such as device type. We extract them from turns and refer to them as session-level features. Thus, at a high level, our model contains features of five types: turn-level textual features, turn-level categorical features, turn-level numerical features, session-level categorical features and session-level numerical features.

\begin{figure}[h]
	\begin{center}
		\includegraphics[width=1.0\textwidth]{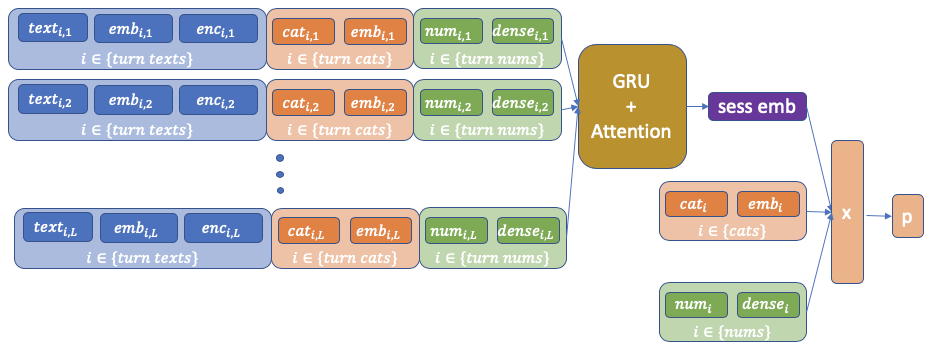}
	\end{center}
	\caption{Overall architecture of our deep neural model for user satisfaction prediction.}
	\label{fig:submodel_architecture}
\end{figure}

Taking these features as input, our model employs a deep-wide style network~\cite{cheng2016wide} to accommodate both structured and unstructured input with a multi-layered structure to perform encoding at both turn and session level as shown in Figure~\ref{fig:submodel_architecture}.

\noindent For each turn, we process user input text and Alexa response text separately by first converting them to word embeddings and then applying a GRU layer followed by an Attention layer to encode a sequence of wording embeddings into a high dimensional sentence vector. We then represent the turn by concatenating the sentence vectors of user input, Alexa response, the turn-level embeddings of categorical features, and the encoding of numerical features. For numerical feature encoding, we pass numerical features into a nonlinear dense layer to map them into a high dimensional vector.
Next, we encode a sequence of turn-level encodings with another GRU and Attention layers and concatenate the resulting vector with the encoding of session-level numerical features and the embeddings of session-level categorical features.
Finally, we pass the session encoding through two nonlinear dense layers followed by a \textit{sigmoid} layer to produce prediction $p$. 

With our deep neural architecture, we build two user satisfaction predictors: 1) a user feedback prediction model (FP) trained on user feedback data; 2) a fallback prediction model (HP) trained on human annotation data.
Note that to make the user feedback prediction model more generalizable, we remove the feedback prompt and answering turn from the session data for both training and evaluation.

\subsection{Hybrid approach}\label{fusion_layer}
\begin{figure}[h]
	\begin{center}
		\includegraphics[width=0.5\textwidth]{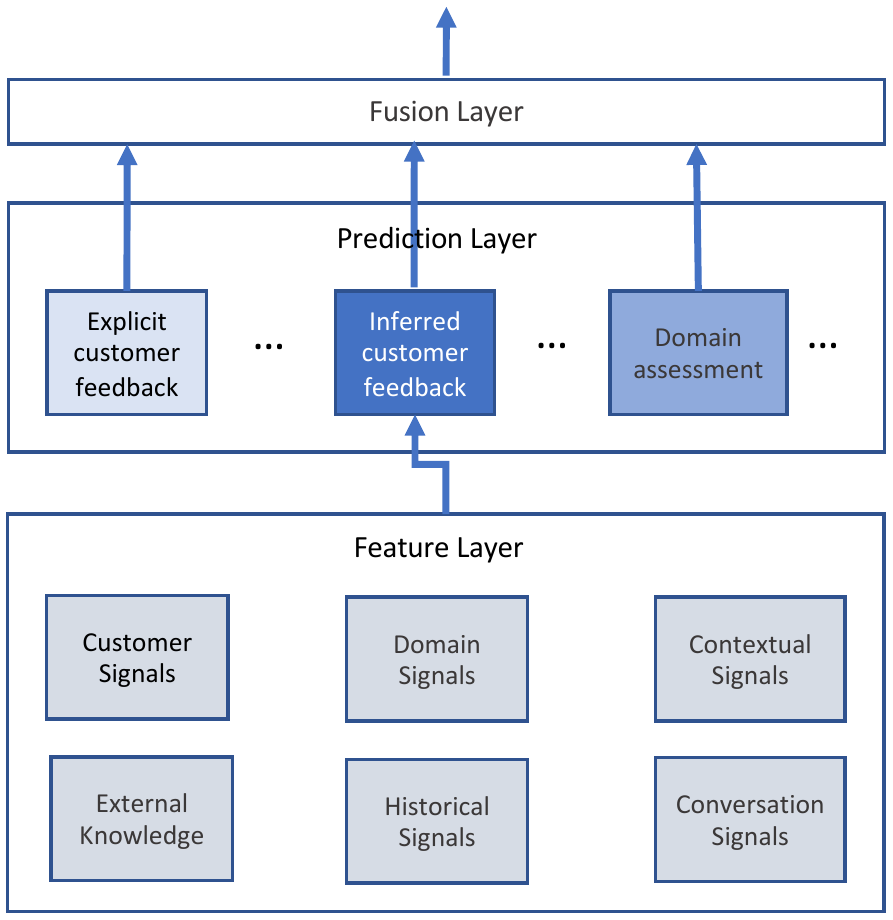}
	\end{center}
	\caption{Overall architecture of our hybrid model.}
	\label{fig:hybrid_architecture}
\end{figure}

The goal of our hybrid approach is to accurately predict user satisfaction by fusing different types of candidate inputs for assessing user satisfaction, as depicted in Figure~\ref{fig:hybrid_architecture}. Roughly, there are three different candidate inputs that are captured in the prediction layer, such as explicit user feedback, inferred user satisfaction, and skill-provided assessment:
\begin{itemize}
\item \textbf{Explicit user feedback} Since user satisfaction is not directly observable, we ask user for post-experience feedback as the best proxy. This is the most direct method of determining whether the experience was satisfying; however, there are some gaps: 1) frequent feedback request introduces friction, thus we should cautiously use it with a controlled rate; 2) it is biased toward positive feedback as we don’t have an opportunity to collect feedback when there are barge-in and early termination that are strong indicators of negative user experience; 3) its coverage is currently limited to a set of whitelisted experiences. 
\item \textbf{Inferred user satisfaction} User feedback is not always available. Predictive model allows us to measure user satisfaction even when user feedback is not directly collected. To build accurate machine-learned predictors, in the feature layer, we consider various input features such as conversation history, contextual features, domain signals, user profile, historical features and external knowledge. Specifically, we utilize the FP and HP models previously described in Section~\ref{dnn_model}.
\item \textbf{Skill-provided assessment} While there are several implicit indicators of user dissatisfaction that are skill agnostic, assessing positive user satisfaction often requires knowledge of the target skill, and access to skill-specific signals. For example, in the media consumption domain, 30-sec playback is commonly used as implicit signal of user satisfaction. In a map/navigation application, we may declare success if we see no changes to the destination or route cancellation within a certain time interval (e.g., 15 secs). In a ticket-booking application, we may use a booking confirmation signal, followed by absence of cancellation within a certain time interval (e.g., 5 mins). Incorporation of skill-provided assessment is out of scope and we leave it for our future work. 
\end{itemize}
As we are at an early stage of leveraging heterogeneous prediction sources, our fusion layer follows a simple waterfall policy to determine whether Alexa’s action/response was satisfying: 
\begin{enumerate}
    \item If explicit user feedback is present and interpretable, then determine user satisfaction accordingly based on the feedback. Otherwise, go to the next step. 
    \item In case, the FP model(the feedback prediction model) is eligible for the experience, predict user satisfaction with the FP model which is trained to predict user feedback. If the prediction confidence is high, then determine user satisfaction accordingly based on the prediction. Otherwise, go to the next step. The confidence threshold is tuned on a separate development dataset.
    \item As fallback, predict user satisfaction with the HP model (the fallback prediction model), which is trained on human annotation to cover all types of experiences.
\end{enumerate}

\section{Experiment}\label{expriment}
In this section, we describe the data sets and experimental setup, and present the experimental results of our hybrid approach.

\subsection{Data sets}
For our experiments, we have whitelisted 43 Alexa intents and collected post experience user feedback at 0.01\% sampling rate. As a result, we have collected 1.3 million data points which is split into training, validation and test set with 
$80\%$, $10\%$, and $10\%$ ratios. The training and validation sets were used to train the FP model and the test set was secured to be used only for reporting. 
For human annotation data, we took a recent chunk of historical Alexa experience annotations that amount to 0.5 million data points. This data set covers all intents and is split into training, validation and test data sets with $80\%$, $10\%$, and $10\%$ ratios.
Note that the size of feedback dataset is larger than that of human annotation dataset as the feedback collection process is much faster and cheaper than the human annotation process.



Due to the limitations in user feedback coverage, we designed an algorithm to build a composite ground-truth test set which weaves user feedback data with human annotation data. The basic idea is we use collected user feedback as ground truth for traffic segments that are eligible for feedback elicitation and use human annotation for all other traffic segments. The ineligible traffic segments include the following:
\begin{itemize}
    \item Intents are not whitelisted for feedback elicitation.
    \item Barge-in or termination request from users. 
    \item Unhandled requests, Alexa saying “I’m sorry...” or no response.
    \item Other feedback types than Yes or No.
\end{itemize}
We compute the proportions of each case in the live traffic and bring in human annotation data for the corresponding amounts. The detailed ground-truth test set construction algorithm is listed in Algorithm~\ref{gt_generation_algorithm}.

\begin{algorithm}[ht]
\caption{Building ground truth dataset}
\label{gt_generation_algorithm}
\SetAlgoLined
\SetKwInOut{Input}{input}
\SetKwInOut{Output}{output}

\Input{$\{N_s\}$ //$N_s$ is the number of ground truth examples for segment $s$ \newline 
$\{H_s\}$ //$H_s$ is a set of human annotated examples for segment $s$; $H^{f}_s$ denotes a subset of examples that could be eligible for user feedback and  $H^{\lnot f}_s$ ineligible\newline
$\{F_s\}$ //$F_s$ is a set of user feedback examples for segment $s$ \newline
$\{R^i_s\}$ //$R^i_s$ is an ineligible rate for user feedback for segment $s$ \newline
$\{R^o_s\}$ //$R^o_s$ is a rate of receiving user feedback categories other than yes/no for segment $s$ \newline
$S$ //$S$ is the entire set of segments \newline
$S_f$ //$S_f$ is a set of segments whitelisted for user feedback
}
\Output{$\{G_s\}$ //$G_s$ is a set of ground truth labels for segment $s$}
\For{$s \gets S$}{
	\eIf{$s$ not in $S_f$}{
		$G_s$ $\gets$ \text{sample}\: $N_s$ \text{examples from} $H_s$\;
	}{
		$n^{hi}_s = R^i_s * N_s$\;
		$n^{ho}_s = R^o_s * (N_s - n^{hi}_s)$\;
		$n^{f}_s = N_s - n^{hi}_s - n^{ho}_s$\;
		$G^{hi}_s$ $\gets$ \text{sample}\: $n^{hi}_s$ \text{examples from} $H^{\lnot f}_s$\;
		$G^{ho}_s$ $\gets$ \text{sample}\: $n^{ho}_s$ \text{examples from} $H^f_s$\;
		$G^{f}_s$ $\gets$ \text{sample}\: $n^{f}_s$ \text{examples from} $F_s$\;
		$G_s = G^{hi}_s \cup G^{ho}_s \cup G^f_s$\; 
	}
}
\end{algorithm}

\subsection{Experiment results}
\begin{figure}[h]
	\begin{center}
		\includegraphics[width=1.0\textwidth]{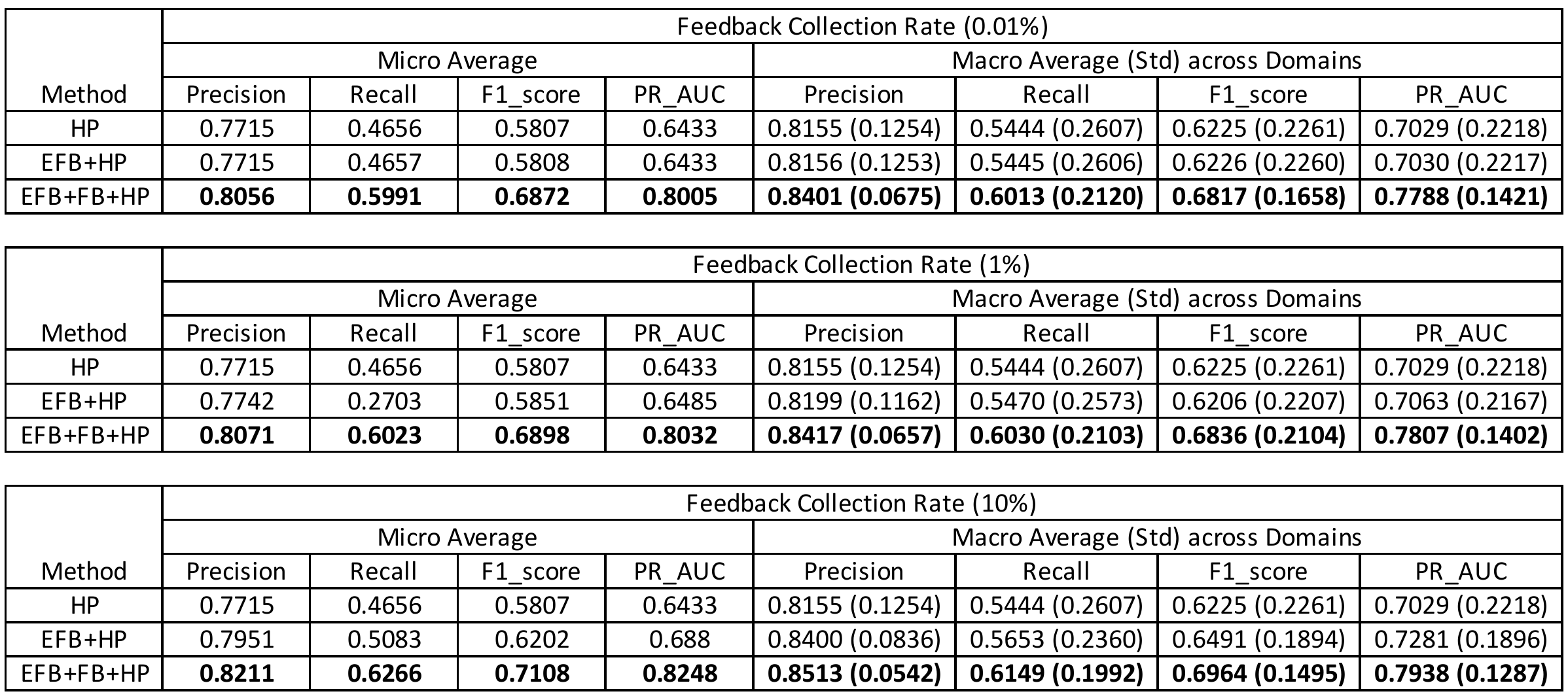}
	\end{center}
    \caption{Evaluation results on the ground-truth test data set with varying feedback collection rates.}
    \label{fig:evaluation_results}
\end{figure}

Our experiment results are presented in Figure~\ref{fig:evaluation_results}. To demonstrate the effectiveness of leveraging user feedback, we compare the following three approaches: (1) \textit{HP}: solely relying on the HP model, (2) \textit{EFB + HP}: fusion of explicit user feedback and HP model prediction according to our hybrid approach and (3) \textit{EFB + FP + HP}: fusion of explicit user feedback, FP model prediction and HP model prediction according to our hybrid approach. 
Note that whenever users provide explicit user feedback, our hybrid approach takes it as output instead of making any inference (based on the fusion approach described above), meaning whenever a user feedback is explicitly given, our hybrid approach can trivially make the right prediction. Thus, evaluating our hybrid approach requires a parameter that controls the rate at which we assume users provide explicit user feedback. Specifically, given a feedback collection rate, we mark user feedback instances as “given by user” in the ground-truth test data until the rate is met. Then for the marked instances, our hybrid approach takes the associated user feedback as its prediction.
In our experiments, we varied feedback collection rate to have a value among the following: $0.01\%$, $1\%$, and $10\%$. $0.01\%$ is the feedback collection rate we chose to use for the experiment, and the other two rates are hypothetical for projective purposes. 

\noindent The micro-averaged results at feedback collection rate of $0.01\%$ clearly demonstrate a large gain that our proposed hybrid approach (i.e. \textit{EFB + FP + HP}) brings in. Compared to a conventional approach (i.e. \textit{HP}, a prediction model trained on human annotation data), precision, recall, F1-score, and PR-AUC (Precision-recall area under curve) metrics are improved by $4.4\%$, $28.7\%$, $18.3\%$, and $24.4\%$, respectively. Looking at the results of \textit{HP} and \textit{EFB + HP}, it is worth mentioning that explicit user feedback barely moves metrics at our current feedback collection rate as the small amount of collected user feedback is easily diluted by the enormous amount of traffic covered by the HP model. This, in turn, signifies the generalization power that the FP model offers, beyond the sparse user feedback samples, enabling us to make accurate predictions for those experiences that are eligible for feedback elicitation but not triggered for elicitation. 


The proposed hybrid model also outperforms the other approaches in macro-averaged results, as shown in the \textit{Macro Average} section of the table. \textit{Macro Average} means that we perform two-staged averaging where we first calculate micro-averaged metrics per each domain, then take a simple average over all the domains. To avoid futile distortion in macro-averaged results due to long tails, we selected top 20 domains that covered $\sim98\%$ of the test set. The smallest standard deviation of the proposed hybrid approach indicates that our approach predicts user satisfaction in a more consistent manner across domains than the other approaches which is a critical property to allocate fair amounts of traffic to each domain according to their service quality. 

With two hypothetical feedback collection rates at $1\%$ and $10\%$, one can clearly see how the increased amount of explicit user feedback impacts the accuracy of our hybrid approach, as shown in the middle and bottom tables. As expected, as we collect more feedback, our hybrid approach makes more accurate predictions and performs in a more consistent fashion across all the domains. Although a blind increase of feedback elicitation rate can cause significant friction in user experience, higher feedback elicitation rates can be safely applied to some targeted segments of traffic without the risk of incurring bad user experience.   


\section{Related work}\label{related_work}

One of the conventional approaches to evaluate the quality of intelligent assistant systems is to measure the relevancy of the response of the system using some IR measures such as precision and cumulative gain measures such as NDCG~\cite{Javelin:81,Saracevic:88}. This approach, however, requires human judgement for the relevancy measures, which is generally costly and hard to scale. Such relevancy-based metrics, however, often do not capture the holistic view of system performance such as user satisfaction. To overcome this, some prior works in the search domain have studied users' behavioral patterns to infer their satisfaction level with respect to search results~\cite{Ageev:13,Jiang:15,Hassan:13,Kim:13}. There is also an attempt to understand the relationship between search engine effectiveness and user satisfaction~\cite{Al-Maskari:07}.

In the area of spoken dialogue system, PARADISE~\cite{walker-etal-1997-paradise} proposed a
framework for evaluating spoken dialogue, specifying the relative contribution of various factors via a linear regression model. For modern intelligent assistants, there was an in-depth study about user satisfaction by classifying the user-system interaction patterns into several categories~\cite{Kiseleva:16}. Another work proposed a research agenda about context-aware user satisfaction estimation for mobile interactions using gesture-based signals~\cite{Kiseleva:17}. These work~\cite{spyros_neurIPS:19}\cite{spyros_sigdal:19} estimated conversation quality via user satisfaction estimation. However, most prior works are annotation intensive. There is an interesting work that pointed out the issues with annotation-based approaches~\cite{Aroyo:15}, which aligns with our motivation toward feedback-based user satisfaction estimation approaches, and even further with our hybrid approach. The ability to accurately predict user satisfaction enables a conversational agent to evolve in a self-learning manner. This overview article~\cite{Sarikaya_PDA:17} about personal digital assistants discussed about user experience prediction using customer feedback. A recent work on Alexa showed how Alexa learns to fix speech recognition and language understanding errors by leveraging an automated user satisfaction predictor without requiring any human supervision~\cite{Ponnusamy:20}.

\section{Conclusion}
\label{conclusion}
In this work, we proposed an effective hybrid approach that outperforms conventional approaches that are solely based on human annotation in the user satisfaction prediction problem. We started from the limitations of the approaches based on human annotation, were motivated to utilize direct user feedback that is not only more direct in capturing user satisfaction, but also more scalable and cost-effective. Our hybrid approach fuses explicit user feedback, user satisfaction predictions inferred by two machine-learned models, one trained on user feedback data and the other human annotation data via a waterfall policy, resulting in significant improvements in performance metrics. The hybrid model also achieved the most consistent performance across the domains, which is another strength. Our proposed approach has been verified with Alexa, and we believe the approach can be extended to other conversational system and text-based chatbot applications. We will extend the fusion layer of our hybrid approach by leveraging machine learning methods. 


\bibliographystyle{coling}
\bibliography{metrics_fusion}

\end{document}